\documentclass[aps,prb,reprint,superscriptaddress,amsmath,amssymb, aps]{revtex4-1}
\pdfoutput=1
\usepackage{bm}
\usepackage{graphicx}

\begin{document}

\title{Local Density-of-States Mapping in Photonic Crystal Resonators by Deterministically Positioned Germanium Quantum Dots}

\author{Magdalena Schatzl}
\author{Florian Hackl}
\author{Martin Glaser}
\author{Moritz Brehm}
\author{Patrick Rauter}
\affiliation{Institute of Semiconductor and Solid State Physics, Johannes Kepler University Linz, Altenbergerstra\ss e 69, 4040 Linz, Austria}
\author{Angelica Simbula}
\author{Matteo Galli}

\affiliation{Dipartimento di Fisica, Universit{\` a}  degli studi di Pavia, via A. Bassi 6, 27100 Pavia, Italy}

\author{Thomas Fromherz}
\email{thomas.fromherz@jku.at}
\author{Friedrich Sch\"affler}

\affiliation{Institute of Semiconductor and Solid State Physics, Johannes Kepler University Linz, Altenbergerstra\ss e 69, 4040 Linz, Austria}
%

\date{\today}

\begin{abstract}
We report on mapping of the local density of states in L3 photonic crystal resonators (PCR) via deterministically positioned single Ge quantum dots (QDs). Perfect site-control of Ge QDs on pre-patterned silicon-on-insulator substrates was exploited to fabricate in one processing run almost 300 L3 PCRs containing single QDs in systematically varying positions in the cavities. The alignment precision of the QD emitters was better than 20\,nm. This type of parallel processing is essentially based on standard Si device technologies and is therefore scalable to any number and configuration of PCR structures. As a first demonstrator, we probed the coupling efficiency of a single Ge QD to the L3 cavity modes as a function of their spatial overlap. The results are in very good agreement with finite-difference time-domain simulations. 
\end{abstract}

\maketitle




\section{Introduction}

Over the last years the prospects of monolithically integrated on-chip optical interconnects in silicon data storage and processing devices \cite{Alduino_2007} have triggered the development of both Si-based optical components \cite{Assefa_2010, Assefa_2011, Moss_2013} and light sources based on group-IV materials \cite{El_Kurdi_2004, Liu_2010, Rong_2005, Xia_2007, Suess_2013, Ondic__2013, Shakoor_2012, Wirths_2015, Grydlik_2016}. Among the latter, self-organized Ge quantum dots (QDs) are the most widely studied type of light emitters. Typically, Ge QDs are embedded in a photonic crystal slab (PCS) or -resonator (PCR) \cite{Xia_2006, Xia_2009, Tsuboi_2012, Hauke_2010, Hauke_2012} to enhance the light emission probability in the $1.3-1.5\,\mu$m telecommunication range via the Purcell effect \cite{Purcell_1946}. In most investigations, randomly nucleated Ge QDs were employed, which evolve in the Stranski-Krastanow growth mode when grown under compressive strain on a Si(001) substrate \cite{EagleshamPRL1990, MoPRL1990}. Meanwhile, site-control of Ge QDs on pit-patterned substrates \cite{ZhongAPL2003, ZhongAPL2004, StoicaN2007} has been developed to an unprecedented degree of perfection \cite{GrydlikN2013}. It was therefore a straightforward step to embed perfectly ordered Ge QDs in a commensurable manner into two-dimensionally periodic PCSs \cite{JannesariOE2014}. We could recently demonstrate that with this approach a specific mode of a two dimensional PCS can be selectively excited by aligning a QD emitter array with the maxima of the local densities of states (LDOS) \cite{JannesariOE2014} of the desired mode. In a similar way, PCRs were implemented, where a single Ge QD emitter was deterministically positioned at 
the very location of the one missing air hole in a PCS that defines an H1 cavity \cite{Schatzl2I1ICoGIPG2015, ZengOE2015}. 
The LDOS is a key quantity for understanding and controlling light-matter interaction in photonic structures \cite{Purcell_1946} and several experimental techniques for LDOS mapping have been developed in recent years. In early attempts, one QD in a randomly nucleated ensemble was singled out, and a dedicated PCR was fabricated around it \cite{BadolatoS2005a}. In this serial approach, mapping requires a separate PCR fabrication step for each QD location in the cavity. More recent techniques employed cavity scanning via nano-emitters that were attached to a scanning probe \cite{FrimmerPRL2011}, by the focused electron beam of a scanning electron microscope (SEM) \cite{SapienzaNM2012}, or by a near field optical microscope \cite{IntontiPRB2008}. In this contribution, we use deterministically positioned single Ge QDs as a means for mapping the LDOS of the resonator modes in PCRs. In an approach suitable for parallel lithographic processing we implemented in a single run a large number of L3 PCRs with different periodicities $a$ of the hexagonal air-hole pattern and varying filling factors $r/a$, where $r$ is the air-hole radius. For each parameter set $\left\{a, r/a \right\}$ ten L3 cavities were fabricated in which the probing Ge QD was systematically moved along the main symmetry axis of the cavity. In this way, we were able to compile line scans of the LDOS in L3 cavities with systematically varying geometry parameters. This novel approach is essentially based on standard Si device technology and, therefore, requires only one alignment step between the two mask layers that define a complete set of resonators and the assigned set of QDs, respectively. Thus, LDOS mapping with deterministically positioned Ge QDs is scalable to arbitrary numbers and configurations of 2D photonic structures.

\section{Sample Layout}
\label{Sample Layout}

The general layout of the L3 PCRs for our experiments is loosely based on the design in Ref. \cite{MinkovSR2014}. We adapted the geometry parameters of the reference design to our experimental constraints, in particular the slab thickness $h$, which is mainly defined by the epilayer thickness, and the air-hole radius $r$, which needs to exceed the maximum QD radius in our technology approach (see next section). Therefore, we set $r\geq75$\,nm and $h=220$\,nm as input parameters of finite-difference time-domain (FDTD) simulations in the open source implementation MEEP \cite{OskooiCPC2010}. The single Ge QD in the cavity was approximated by a cylinder of radius $r = 75$\,nm, a height $z$ between 20\,nm and 50\,nm and a dielectric constant of $\varepsilon$ = 13.56. The chosen $\varepsilon$  value corresponds to the average Ge concentration of $\sim$ 37\,\% in a typical, site-controlled Ge QD deposited at 700$^\circ$C \cite{SchuelliPRL2009, PezzoliNRL2009}. The simulated Ge QD was located at systematically varying positions along the centerline connecting the three missing air holes that define the L3 cavity ($x$-direction in the following). In addition, a small indentation on the slab surface directly above the QD was taken into account, which remains even after high-temperature Si capping (see next section) as a residue of the initially etched pit for QD nucleation \cite{BrehmJoAP2011}. The FDTD results revealed that the small amount of higher index material in the QD leads to a slight shift of the cavity modes to higher wavelengths, whereas missing material ($\varepsilon$ = 1) above the QD causes shifts in opposite direction. Moreover, the respective shifts depend on the location of the QD in the cavity. 

\begin{figure}[t]
\centering
\includegraphics[width=0.8\linewidth]{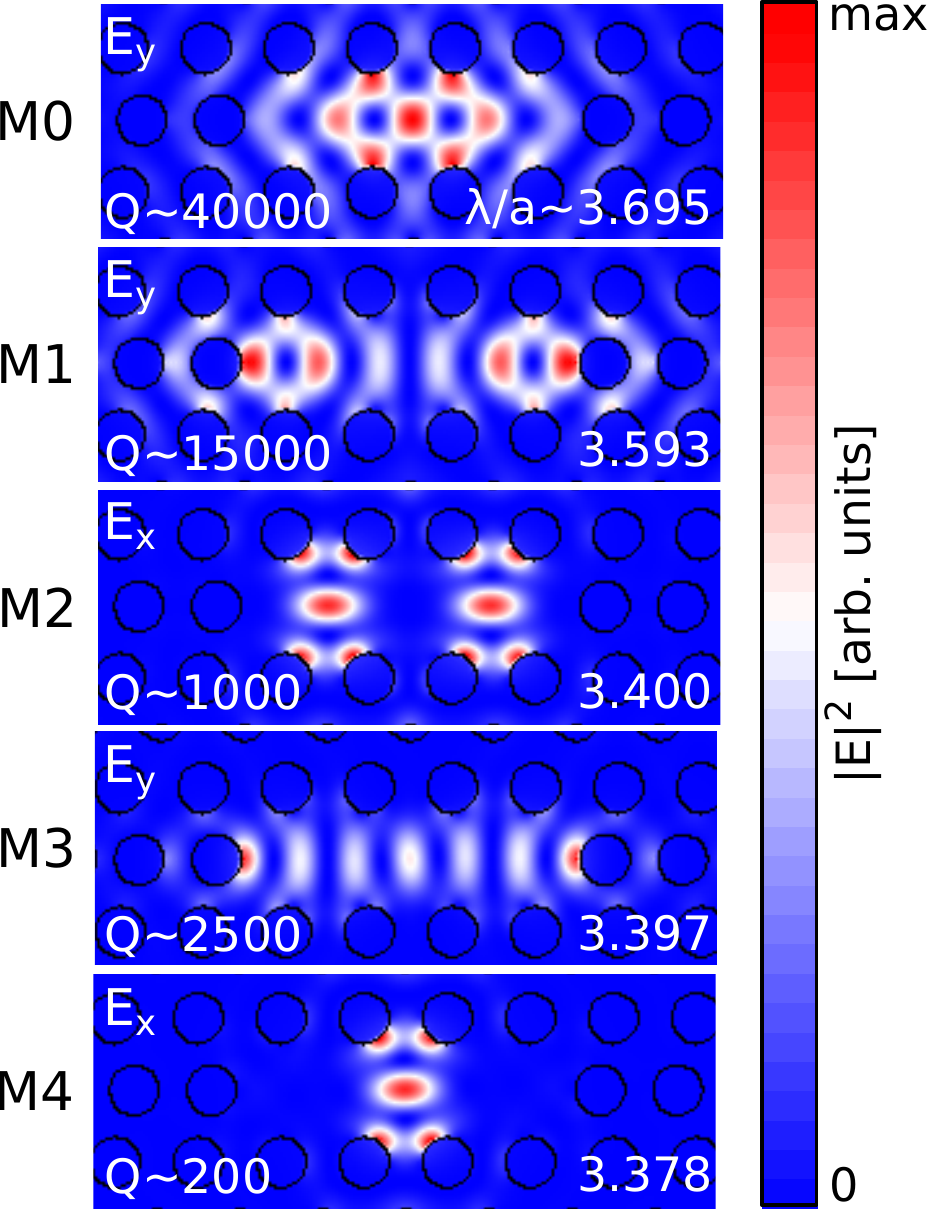}
\caption{Simulated electric-field energy density distributions of five 
 cavity modes of an ideal L3 cavity without Ge QD). The resonator design is optimized for the $Q$ factor of the ground mode M0  as proposed in Ref. \cite{MinkovSR2014}. The labels indicate the dominant electric field polarization of the mode (E$_x$ or E$_y$), the calculated $Q$ factor and the normalized wavelengths in units of the PCS period $a$.}
\label{fig:mode-pattern}
\end{figure}

For an ideal cavity without QD and with planar surfaces, Fig.\,\ref{fig:mode-pattern} shows a compilation of the simulated electric-field energy distributions for the five cavity modes M0-M4. Over the limited wavelength range of the cavity modes the electric-field energy represents to good approximation the LDOS for an electric dipole emitter oriented parallel to the electric field direction at its position \cite{KoenderinkOL2010}. In addition, the dominant in-plane polarization (E$_x$ or E$_y$), the calculated $Q$ factors and the resonance wavelengths in units of $a$ are denoted in the frames of Fig.\,\ref{fig:mode-pattern}. Evidently, each cavity mode shows a unique LDOS pattern, which makes L3 cavities particularly well suited for mapping with single QD emitters positioned at systematically varying locations in the cavity. For the investigated sets of geometry parameters the simulations resulted in a group of higher $Q$ factors (modes M0 and M1) with values of 42000 and 15000, respectively, and a group of low-$Q$ modes (M2-M4) with values below 2000 (Fig.\,\ref{fig:mode-pattern}). 

Based on the simulated mode patterns, we designed a pair of mask layouts containing 280 complete PCRs. For these, the periods $a$ are varied in seven steps between 330\,nm and 420\,nm, and each period is implemented with four different hole radii to cover a range of filling factors $r/a$ from 0.305 and 0.325. Variations of both $a$ and $r/a$ shift the energetic positions of the cavity modes \cite{joannopoulos2011photonic}, and thus allow for an adjustment of the spectral overlap between the cavity modes and the emission range of the single QD in each cavity. In each of these geometrically defined PCR sets the QD position is systematically moved in nine steps throughout one half of the L3 cavity along the x-direction, and, as a reference, one cavity in each series contains only the Ge wetting layer (WL), but no QD. 

\section{Sample Fabrication}
\label{Sample Fabrication}

\begin{figure}
\centering
\includegraphics[width=0.8\linewidth]{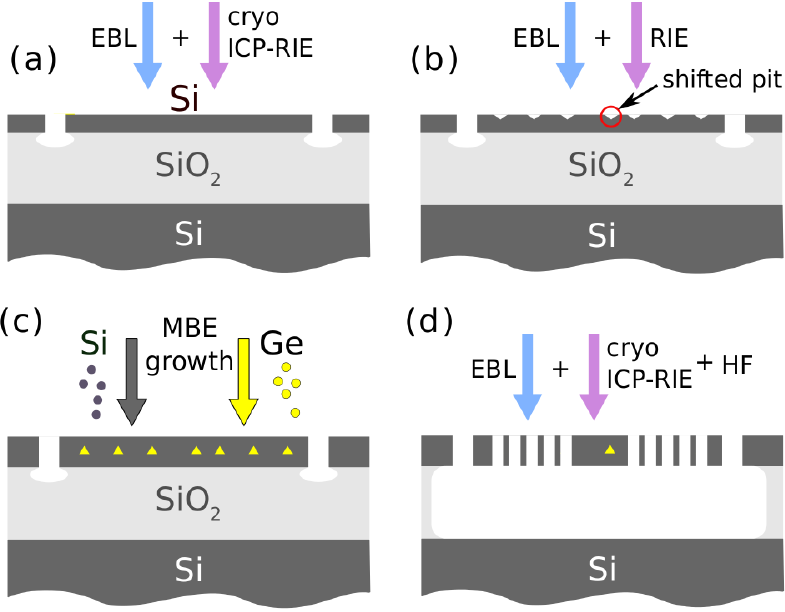}
\caption{Fabrication process for PCRs with a single Ge QD at defined positions within the cavity. (a) Alignment markers defined by EBL are etched deep into the BOX by ICP-RIE. (b) Definition of the
preferential nucleation sites for Ge QDs. (c) MBE growth of the QDs and the Si cap
layer. (d) Fabrication of the PCR by ICP-RIE and subsequent BOX removal. In this step all QDs in registry with the air holes of the PCR are removed.}
\label{fig:fabrication}
\end{figure}

A serious concern for optical investigations of SiGe QDs on silicon-on-insulator (SOI) substrates was raised by the recent discovery \cite{Hauke_2012, LoSavioAPL2011} of optically active defects in industrial SOI substrates that emit at room temperature in the same wavelength range as SiGe QDs ($1.3-1.6\,\mu$m). Meaningful optical experiments with SiGe QDs require substrates that are free of such defects. For this work, we acquired custom-made SOI substrates with a 2\,$\mu$m thick buried oxide layer (BOX) on which the manufacturer SOITEC performed an additional thinning/re-bonding/annealing process sequence \cite{MalevilleMSaEB1997}. In this way, remnants of the hydrogen implantation for the smart-cut\textsuperscript{\textregistered} process \cite{MalevilleMSaEB1997} should have been removed in a similar way as reported in Ref. \cite{Hauke_2012}. Moreover, the additional thinning process led to a device layer thickness of just 70\,nm, which allowed us to vertically center the Ge QDs in the designed slab thickness of 220\,nm (see Fig.\,\ref{fig:fabrication}(c)). To check the efficient removal of fabrication-induced optically active defects, we performed photoluminescence (PL) experiments on both the bare SOI substrates and processed reference samples. The latter underwent all subsequently described fabrication steps for L3 cavities with the exception of the Ge epilayer. Thus, the reference cavities contained all defects that were either caused by SOI fabrication or induced by our subsequent PCR processing sequence. In the reference PCR we found only negligible indications of defect PL in the spectral range of the Ge QDs, which are $\sim 500$ times smaller than the QD related signals  (see Fig.\,\ref{fig:4}(d) further down). We can therefore be sure that the PL signals of the PCR structures in this work originate from the deposited Ge epilayers.

The fabrication of L3 PCRs followed a modified version of the processing sequence which we have developed recently for the deterministic embedding of site-controlled Ge QDs into two-dimensional photonic crystals and H1 PCRs\cite{JannesariOE2014,Schatzl2I1ICoGIPG2015}. 
The four basic processing steps are depicted in Fig.\,\ref{fig:fabrication}. In the first step (Fig.\,\ref{fig:fabrication}(a)), alignment marks are etched deep into the BOX of the SOI substrate. They are defined by electron beam lithography (EBL) in a Raith eLine Plus facility equipped with a laser stage that allows for an alignment precision in the 10\,nm range. Anisotropic etching into the BOX is performed at cryogenic temperatures in an inductively coupled (ICP) Oxford 100 reactive ion etcher (RIE) employing an SF$_6$/O$_2$ process \cite{JannesariOE2014}. 

\begin{figure}[t]
\centering
\includegraphics[width=0.8\linewidth]{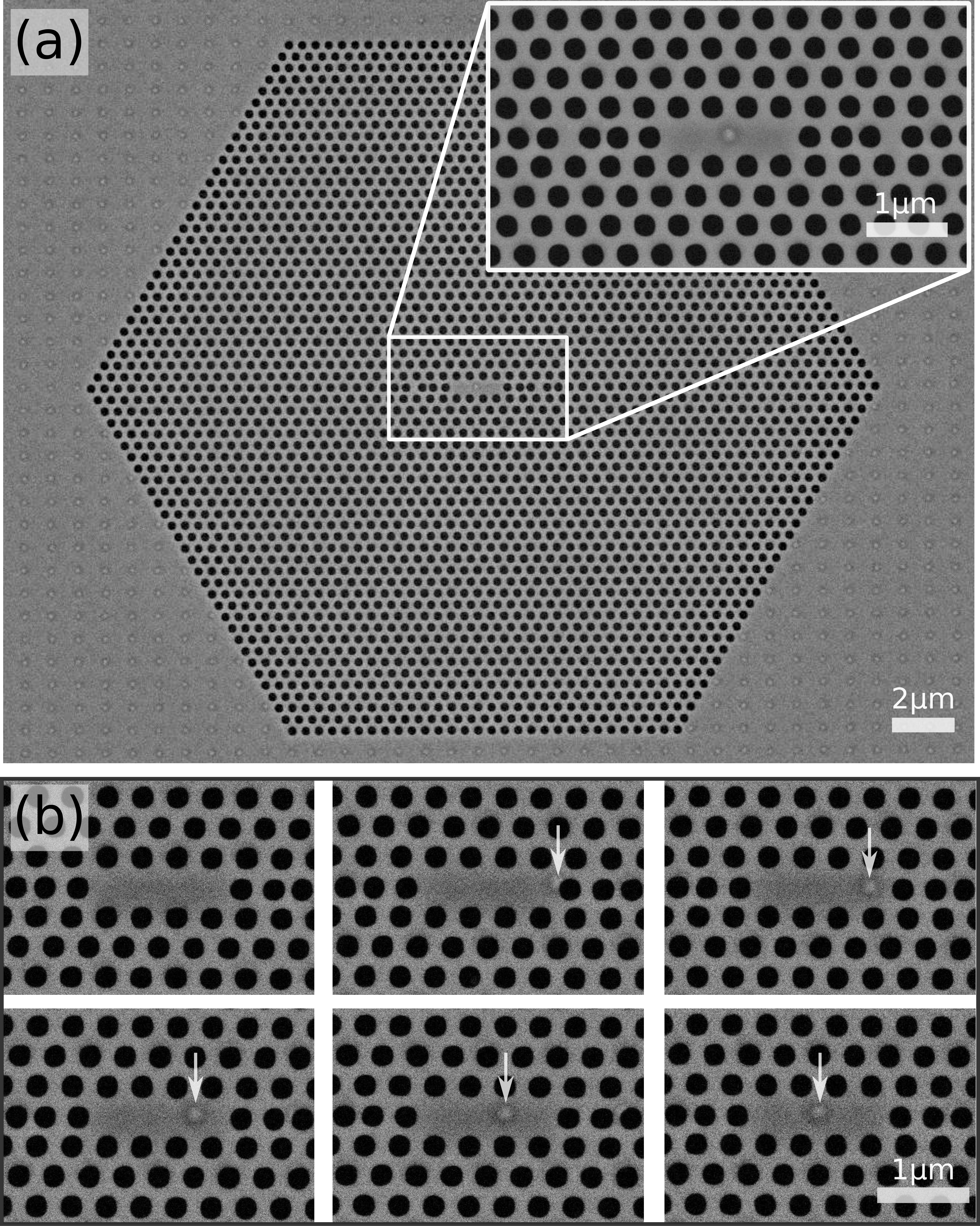}
\caption{SEM images of L3 cavities fabricated during one growth and processing run with a
conformal Si cap to allow imaging of the buried Ge QDs. (a) Complete PCR structure with a
single Ge QD in the center of the L3 cavity (inset). The periodic QD array with twice the
period of the photonic crystal array remains only outside the photonic structure. The inset
reveals the modifications of the air hole positions adjacent to the cavity of our adapted high-$Q$
design. (b) Representative set of six simultaneously fabricated L3 cavities in which the Ge QD (marked by an arrow)
in the cavity was moved along the horizontal center line of the cavity. As a reference, one of these cavities was written without a pit for QD nucleation, i.e. it
contained only the Ge wetting layer (first frame in (b)).}
\label{fig:SEM}
\end{figure}

In the second step (Fig.\,\ref{fig:fabrication}(b)), a set of pits for the preferential nucleation of QDs \cite{ZhongAPL2003, GrydlikN2013, HacklN2011} is fabricated by EBL and a shallow (35$\pm$5\,nm deep) ICP-RIE process. Complete arrays of pits with twice the periodicity of the air hole patterns for the PCRs are realized in this lithography level. Only the one pit that defines the QD position in each of the subsequently processed cavities is slightly moved out of registry in a series of ten adjacent PCR structures. In this way, the cavity QD is shifted systematically between the center and the terminating air hole at one edge of the of the L3 cavity (Fig.\,\ref{fig:SEM}(b)). The alignment marks are used for positioning of the pit-pattern mask, thus defining a reference frame that will be employed again for the later PCR fabrication step. Writing complete pit arrays with comparably small periods is essential for tight size control and perfect ordering of the QDs\cite{GrydlikN2013}.

In the third processing step (Fig.\,\ref{fig:fabrication}(c)) the epitaxial layers are deposited. The samples are prepared for epitaxy by resist removal in organic solvents and an oxygen-plasma followed by an RCA cleaning procedure \cite{KernJES1990}. A final treatment in 1\% hydrofluoric acid (HF) removes the natural oxide and provides a hydrogen-terminated surface that stabilizes the pit pattern during subsequent heat treatments \cite{LichtenbergerAPL2003}. The templates are then immediately transferred into the load-lock chamber of a Riber SIVA 45 molecular beam epitaxy (MBE) system and subsequently degassed in the ultra-high vacuum environment of the growth chamber at 750$^\circ$C for 45 min. The epitaxial layer sequence consists of a 40\,nm thick Si buffer, 0.66\,nm of Ge and a 110\,nm thick Si cap. During buffer growth the substrate temperature is ramped from 450 - 550$^\circ$C and is then kept constant at 700$^\circ$C for Ge deposition. Ge growth leads to an approximately three monolayer (ML) thick, Ge-rich WL that covers the entire sample \cite{BrehmPRB2009}, and arrays of perfectly ordered Ge QDs which nucleate at the centers of the predefined pits \cite{GrydlikN2013} (Fig.\,\ref{fig:SEM}(a)). 

Two versions of the Si capping layer were grown on a pair of samples with otherwise identical growth and layout parameters. For the first one, the substrate temperature was ramped from 350 to 450$^\circ$C to achieve conformal covering of the Ge QDs \cite{ZhangSST2010}. The second cap version commenced again at 350$^\circ$C to preserve the QD shape, but was then ramped to 700$^\circ$C to partly planarize the epilayer surface \cite{BrehmJoAP2011}. The conformally capped samples were mainly used for visualizing the QD positions by SEM (Fig.\,\ref{fig:SEM}). These samples allowed us to validate an overall alignment precision of better than 20\,nm for the complete process. The partly planarized samples were used for optical characterization because their more homogeneous slab thickness leads to better defined cavities. 

In the last processing step (Fig.\,\ref{fig:fabrication}(d)) the initially defined alignment marks are exploited again to align the air-hole mask of the PCRs in such a way that the QDs become centered in the air holes. In this way, all QDs are removed during air hole etching except the one remaining in the cavity. The air holes are ICP-RIE etched with perpendicular sidewalls down to the BOX in an SF$_6$/O$_2$ cryo-process. In a last step, the BOX beneath the PCR structures is selectively removed with 40\% HF to achieve free-standing PCR slabs. 

\section{Optical Characterization}
\label{Optical Characterization}

\begin{figure}[t]
\centering
\includegraphics[width=1\linewidth]{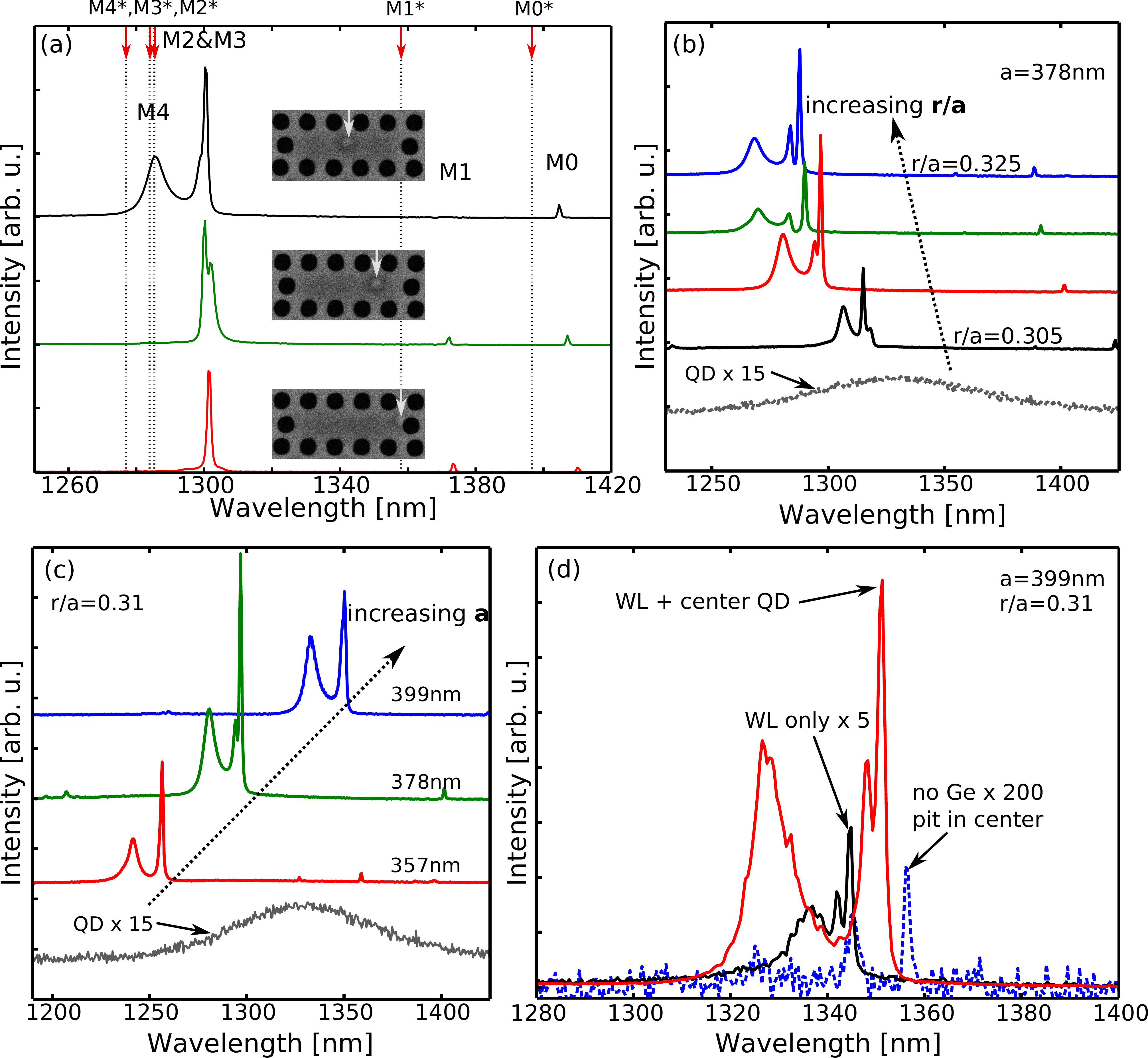}
\caption{(a) PL spectra of an L3 cavity with $a$ = 378\,nm, $r/a$ = 0.31 covering the range of all five modes. Spectra for three different locations of the Ge QD are displayed: center position (black line), edge position (red line) and intermediate position (green line).  The vertical lines and the arrows mark the simulated mode maxima (labeled M0$^*-$M4$^*$), which coincide to within 15\,nm with the associated PL peaks. Coupling of the QD in the cavity to the five modes is strongly position dependent and different for each mode. (b) Shift of the PL spectra with filling factor $r/a$ at fixed $a$ = 378\,nm, and, (c), with period $a$ at fixed $a/r$ = 0.31. In (b) and (c) the QD is at the cavity's center. As a reference,  PL spectra (amplified by a factor of 15) of ensembles of ordered QDs outside the photonic structures are shown in gray. (d) Comparison of PL spectra from a cavity ($a$ = 399\,nm, $r/a$ = 0.31) with the QD in its center (red line) and an identical cavity that contains just the WL but no QD (black line, scaled up $\times 5$). As a reference, a PL spectrum from an all-Si PCR  is shown that has seen all processing steps with exception of the Ge epilayer (blue line,  scaled up $\times$ 200).}
\label{fig:4}
\end{figure}

The $Q$ values of the processed cavities were determined by resonant laser scattering experiments, as described in Ref. \cite{GalliAPL2009}. 
 On the Ge-free reference sample, we found $Q$ = 40000 for mode M0, in excellent agreement with the simulation value for this mode. With the Ge epilayer added, the $Q$-factor of this mode decreased to 8200. Although the $Q$ values are not important for the investigations in this contribution, degradation caused by the adding of Ge epilayers can be a concern for other applications. A possible cause could be enhanced roughness of the air hole surfaces introduced by the different etching rates for Si and Ge in the utilized RIE process. To test the basic limitations of our technology sequence, we fabricated and measured also another reference cavity based on the novel high-$Q$ design proposed in Ref. \cite{AlpeggianiAPL2015}. This cavity was implemented on a bare SOI substrate without any epitaxial layer and resulted in a measured $Q$\,=\,1,200,000\,\cite{Simbulatbp2016}, a value that can be considered state-of-the-art. 

Photoluminescence measurements were performed with a confocal micro-PL setup \cite{JannesariOE2014, HacklN2011} consisting of an excitation diode laser operated at 442\,nm and a microscope objective with a numerical aperture of $NA = 0.7$. The latter is used both for laser focusing and for collecting the PL signal from the sample. The detection spot on the sample with a diameter of $\sim 3\,\mu$m is centered at the selected L3 cavity.  For polarization-dependent measurements a $\lambda/2$ plate and a linear polarizer  are used to record spectra in 5$^\circ$ rotation steps of the former. The signal is collected perpendicular to the sample surface with an  acceptance cone covering a solid angle of $0.15\times4\pi$\,sr. The sample is mounted in a cryostat and maintained at a  temperature of 10K. The signal is dispersed in a grating spectrometer and recorded with a liquid-nitrogen cooled InGaAs line detector. To enhance spectral overlap between the cavity modes and QD emission, a rather high laser excitation power of 750\,$\mu$W was chosen which has been shown to cause significant spectral broadening of the time-averaged emission spectrum of single Ge QDs \cite{GrydlikAPL2015}.

Figure\,\ref{fig:4} shows PL spectra over the wavelength range of all five cavity modes. In Fig.\,\ref{fig:4}(a), three spectra are depicted for QD positions at the center, the edge and an intermediate position in the L3 cavity, as highlighted in the inserts. The spectra were recorded on PCRs with $a$ = 378\,nm and $r/a$ = 0.31. Evidently, the PL intensities of the modes depend strongly on the QD position in the cavity. The vertical dashed lines in Fig.\,\ref{fig:4}(a) indicate the simulated peak positions of the five modes (labeled at the red arrows), which are consistently shifted by about 15\,nm toward shorter wavelength with respect to the measured resonances. Such small shifts are indicative for minor deviations between the simulated and the implemented geometry parameters of the PCR. In particular, the (effective) hole radii may be affected by the fabrication process due to, e.g., surface roughness or slight tapering of the air holes. The pronounced influence of small changes of $r/a$ is confirmed by the spectral shifts in Fig.\,\ref{fig:4}(b), where $r/a$ varies between 0.305 and 0.325 at $a$ = 378\,nm, with the QD being in the center position for all four spectra. The influence of the PCR period is shown in Fig.\,\ref{fig:4}(c), where $a$ varies between 336\,nm and 399\,nm at a fixed filling factor $r/a$ = 0.31, with the QD again being located in the cavity center. As expected for PCRs with constant slab thickness $h$, the mode wavelengths shift linearly with $a$ \cite{joannopoulos2011photonic} . Figures\,\ref{fig:4}(b) and \ref{fig:4}(c) also contain a reference signal from an array of bare QDs which was measured on an area outside the PCR structures, where the original QD array is preserved after air-hole etching (Fig.\,\ref{fig:SEM}(a)). 
The reference spectra are multiplied by a factor of 15, highlighting the enhancement of QD emission in the cavities. It is clear from the two data sets in Fig.\,\ref{fig:4}(b) and \ref{fig:4}(c) that within the implemented variation range of the geometry parameters every mode can be spectrally aligned with the intrinsic QD emission band. Also, all five PCR modes show PL peaks in all spectra, i.e. the spectral overlap between the cavity modes and the QD emission band is not very critical under our experimental conditions (see also discussion section).

Figure\,\ref{fig:4}(d) compares PL spectra from a cavity with a centered QD and \{$a$ = 399\,nm, $r/a$ = 0.31\} (red line), and a reference cavity of identical geometry that contains just the Ge WL, but no QD (black line). The latter is scaled up by a factor of five for better visibility. The Ge WL behaves essentially as a quantum well (QW) for holes \cite{BrehmPRB2009} that is evenly distributed over the whole cavity. The QW evidently contributes to the PL signal of the cavity modes, but the signal is by a factor of 5 - 20 weaker than the signal caused by a QD in the cavity. In addition, any features in the PL spectrum of the PCR that depend on the QD's position have to be related to the QD, i.e., they cannot be induced by the delocalized  WL. For completeness, we also show in Fig.\,\ref{fig:4}(d) the PL spectrum measured for a PCR that contains no Ge at all (blue broken line, amplified 200 times) but was otherwise processed identically as the one containing the WL and a single QD (red line in this figure). Evidently, the emission of residual defects is negligibly small  in our PCRs. 

Figures\,\ref{fig:4}(b) and \ref{fig:4}(c) show that the highest-$Q$ mode M0 emits much weaker than modes M2 and M3, even if M0 is spectrally aligned with the maximum of QD emission. This finding is a consequence of the employed PRC design, which was originally developed for a high $Q$ value of M0 and weak out-coupling perpendicular to the PCS \cite{MinkovSR2014}. 

\begin{figure}[t]
\centering
\includegraphics[width=0.9\linewidth]{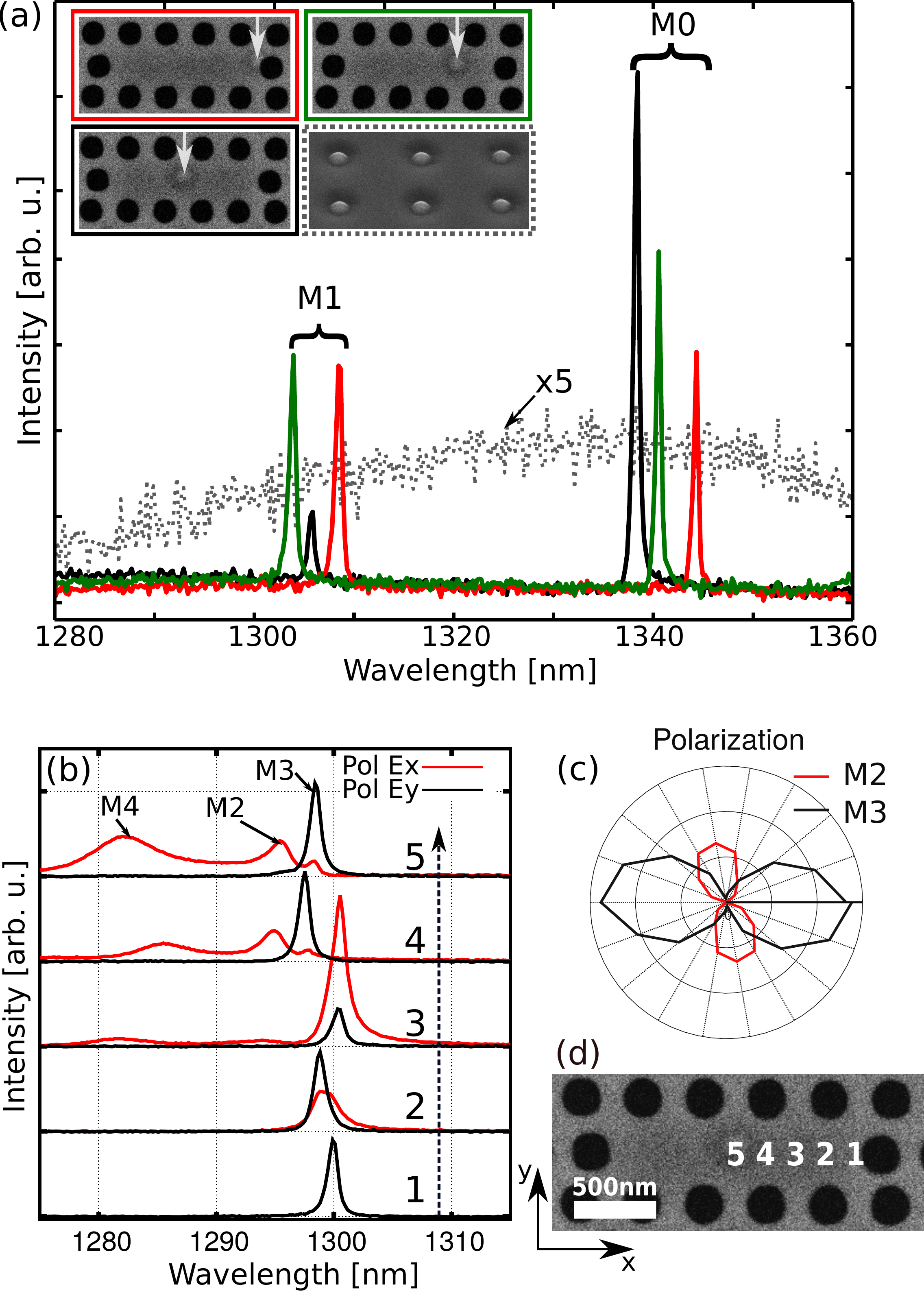}
\caption{(a) PL spectra of cavity modes M0 and M1 for the locations of the QD in the  L3 cavity ($a=336$\,nm, $r/a=0.31$) indicated in the insets (black: center, red:edge, green between center and edge); the reference spectrum (broken line) represents ordered QDs from a region outside the PCRs (see Fig.\,\ref{fig:SEM}(a)) scaled up by a factor of five for better visibility. (b) Modes M2-M3 for the five QD positions 1-5 indicated in (d) in a L3 cavity ($a=357$\,nm, $r/a=0.31$); the two orthogonal in-plane polarizations labeled x and y are displayed, where x corresponds to the long axis of the L3 cavity. (c) The orthogonal polarization behavior of modes M2 and M3 are shown. The different PCS periods for (a) and (b) were chosen to align the cavity modes to the wavelength range of the QD emission band. }
\label{fig:5}
\end{figure}

For a more detailed investigation of the cavity modes and, in particular, their dependence on the QD position in the cavity, we coarsely aligned the spectral overlaps of either the two high-$Q$ modes M0 and M1, or the group of lower-$Q$ modes M2 - M4, by selecting cavities with appropriate geometry parameters. Figure\,\ref{fig:5}(a) shows in the same scale PL spectra containing modes M0 and M1, again for the center-, edge- and intermediate QD positions. The PCS parameters for the three spectra are $a = 336$\, nm and $r/a = 0.31$. The reference signal from a QD array outside the cavities is scaled up by a factor of five. The spectra show two distinct effects of the QD position: For one, the resonance peaks move slightly with the location of the QD in the cavity. These shifts are in quantitative agreement with our simulations if the local modifications of the dielectric constant by the QD itself and the pit-related material depletion in the Si cap above the QD are taken into account. Secondly, the peak intensities vary strongly with the QD location in a manner that is distinctively different for the two modes. The PL intensity of mode M0 reaches its maximum when the QD is in the center position, whereas the M1 signal becomes minimal under these conditions, in qualitative agreement with the calculated LDOS patterns in Fig.\,\ref{fig:mode-pattern}. 

Figure\,\ref{fig:5}(b) shows PL spectra of the modes M2 - M4 for in-plane polarizations in $x$ and $y$ direction. To stay near the maximum of the QD emission signal, the set of PCRs with $a = 378$\,nm was chosen here, again with $r/a$ = 0.31. The five frames in Fig.\,\ref{fig:5}(b) represent spectra for the five equidistantly spaced QD positions indicated in Fig.\,\ref{fig:5}(d). Again, a strong modulation of the peak intensities with QD position is observed, as well as an evident mode polarization. In agreement with the simulations in Fig.\,\ref{fig:mode-pattern}, modes M2 and M4  are orthogonally polarized with respect to mode M3.  In Fig.\,\ref{fig:5}(c), the measured polarization dependent PL intensity is shown in detail for M2 and M3. 

\begin{figure}[t]
\centering
\fbox{\includegraphics[width=\linewidth]{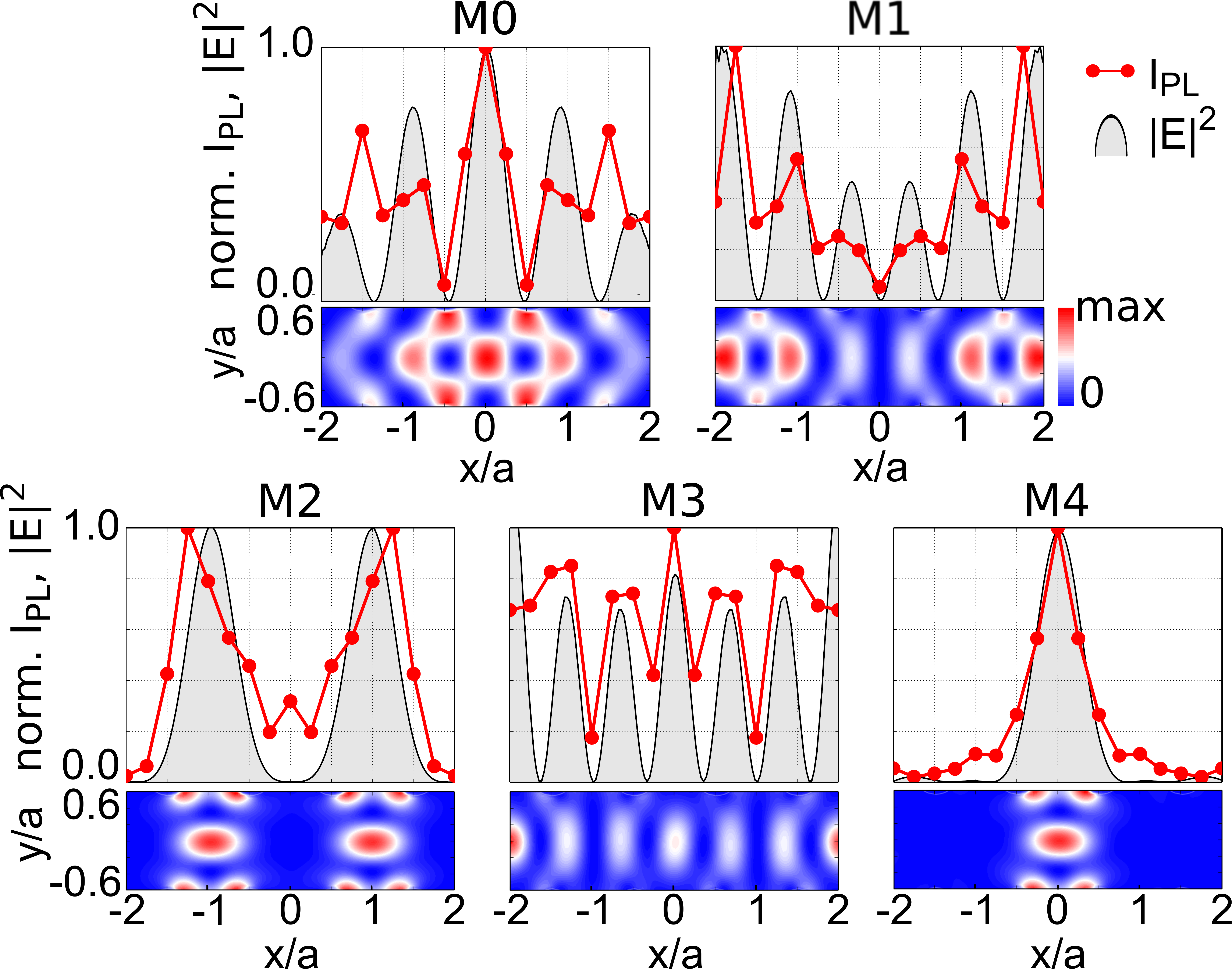}}
\caption{Integrated PL intensities $I_{PL}$ (red connected dots) for QDs positioned equidistantly over an L3 cavity in comparison with electric-field energy  $|\bm E|^2$ line scans through the center of the modes in Fig.\,\ref{fig:mode-pattern} (gray areas). Upper (lower) row: high- (low-) $Q$ modes M0, M1 (M2-M4).  As in Fig.\,\ref{fig:5}, the PL intensities shown  in the upper (lower) row were recorded on cavities with a PCS period of $a$ = 336\,nm (357\,nm) and  $r/a$ = 0.31. The electric field energy maps below each frame are extracted from Fig.\,\ref{fig:mode-pattern} and cover the L3 cavity up to the edges of the first air holes in both $x$ and $y$ directions. Only for PCRs with QDs positioned on the positive $x$ axis PL experiments have been performed. To be able to display the complete cavity, the experimental values were mirrored onto  the frames' negative $x$ axis. All traces are normalized to their respective maximum.}
\label{fig:6}
\end{figure}

Figure\,\ref{fig:6} shows for the five cavity modes complete sets of PL data from all nine QD positions in the cavity in comparison with simulated, normalized LDOS line scans through the cavity center.  Within each panel, the shown data points (connected red dots) were measured with a constant excitation laser power on cavities with identical geometrical parameters.  The shown PL intensities ($I_{PL}$) are integrated over the resonance linewidth and normalized to the maximum value of $I_{PL}$  in the respective panel. 
 For the groups of high-$Q$ and low-$Q$ modes the same geometry parameters were employed as in Figs.\,\ref{fig:5}(a) and \ref{fig:5}(b), respectively. To highlight the mode symmetries, cross sections of the complete L3 cavity are depicted. Since the QD positions in the experiments were only varied over one half of each cavity, the experimental data points were mirrored at the cavity center for Fig.\,\ref{fig:6}. 

\section{Discussion}
\label{sec:Disc}

For non-dissipative cavities with modes well separated by their resonance frequency and/or direction of polarization, the LDOS is well approximated by the Purcell factor $F_P$ \cite{Purcell_1946, KoenderinkOL2010}, which is defined as the ratio between the radiative free-decay lifetime $\tau_{free}$ and the cavity radiative lifetime $\tau_c$. In a generalized form $F_P$ depends both on the spatial and the spectral overlap between cavity and emitter according to \cite{GerardJoLT1999} 
\begin{eqnarray}
&&F_P(\bm r,\omega_e,\hat{\bm{e}}_e) =\nonumber \\
&&=\frac{\tau_{free}}{\tau_c}=F_P^{max} \cdot \frac{\kappa^2}{4(\omega_e-\omega_c)^2+\kappa^2} \cdot \frac{|\bm E(\bm r_e)|^2}{|E_{max}|^2} \cdot|\hat{\bm{e}}_e \hat{\bm{e}}_c|^2 .\nonumber \\
&&
\label{equ:FP}
\end{eqnarray}
The first term on the right hand side (r.h.s) is the maximum Purcell factor $F_P^{max}=\frac{3Q(\lambda_c/n)^3}{4\pi^2V_{eff}}$
with the cavity quality factor $Q$, the resonance wavelength $\lambda_c/n$ in a cavity medium with an index of refraction $n$ and the effective mode volume $V_{eff}$. The second term defines the spectral overlap of a dipole emitter of frequency $\omega_e$  with cavity mode $\omega_c$, where $\kappa=Q/\omega_c$ . The third term gives the spatial overlap expressed in terms of the modal electric field strength $\bm E(\bm r_e)$ at the emitter position $\bm r_e$, and the mode's maximum modulus of the electric field $E_{max}$. The fourth term contains the polarization unit vectors of emitter and cavity. 

In general, for solid state emitters radiative and non-radiative decay compete. If  non-radiative decay is characterized by a time constant $\tau_{nr}$, the quantum efficiency of the radiative decay into a specific cavity mode is given by $\eta=\frac{\tau_{nr}}{\tau_{nr}+\tau_c}$. Here, all radiative decay channels other than the radiative decay into the cavity mode have been neglected. In the limit $\tau_{nr} \ll \tau_c$ the quantum efficiency  $\eta \approx \frac{\tau_{nr}}{\tau_c}$ is proportional to $F_P$. Only in this regime it is possible to map $F_P$ and, thus, the LDOS via the PL \emph{intensity} of a set of emitters placed at different positions of the LDOS profile. In addition,  $\tau_{nr}$ needs to stay constant  within that set.   
On the other hand, a direct measurement of $\tau_c$ and, thus, $F_P$ by time dependent PL experiments is not possible in this limit, since the experimentally accessible  decay time $\tau_{tot}$ of $I_{PL}$ is dominated by $\tau_{nr}$ according to  $\tau_{tot}^{-1}=\tau_{nr}^{-1}+\tau_c^{-1}$. Instead, elaborate methods based on the modeled farfield of the PCRs  have to be used to experimentally estimate $F_P$ \cite{SumikuraOE2016}, which are clearly beyond the scope of this work. 

By measurements of $\tau_{tot}$ in a time-correlated single photon counting setup we verified the aforementioned preconditions for LDOS mapping using $I_{PL}$ as probe. 
With the exception of the edge positions we found for all data points in Fig.\,\ref{fig:6} $\tau_{tot}=28 \pm 5$\,ns   \cite{lifetime}. This findings imply both $\tau_{nr} \ll \tau_c$ and  $\tau_{nr}$ being virtually independent of the QD position within the cavity. As a consequence, the overall agreement between the PL experiments and the simulated LDOS line scans in  Fig.\,\ref{fig:6} is remarkably good and clearly demonstrates the successful mapping of the L3 LDOS  in our deterministic QD positioning approach. 

Evidently, our experiments are dominated by the third term on the r.h.s. of Eq.\,\ref{equ:FP}, whereas the second term has only minor influence due to the spectral diffusion of the SiGe QD emission.  
We know from power-dependent PL measurements on single Ge QDs without resonator structure that Ge QDs on Si(001) have rather broad PL signals, even in the limit of very low excitation densities \cite{GrydlikAPL2015}. These broad peaks were attributed to the interplay of several mechanisms that can be related to the type-II band offset and the indirect band gaps of  Si and Ge \cite{GrydlikAPL2015}. The type-II band offset leads to strong hole confinement in the Ge QD, whereas the small intrinsic conduction band offset allows only for weak electron confinement in local strain pockets of the Si matrix \cite{BrehmNJP2009, BrehmNRL2010}. Moreover, the strain fields lift the sixfold conduction band degeneracy of Si, leading to a dense and complex system of energy levels \cite{BrehmNJP2009, BrehmNRL2010}. As a consequence, electron fluctuations are to be expected at the high excitation densities employed here  leading to spectral diffusion and significantly broadened emission spectra in time averaged experiments \cite{GrydlikAPL2015, KlenovskyPRB2012}. 
We want to emphasize that our LDOS mapping method greatly benefits from spectral diffusion. While a pronounced spectral diffusion might be disadvantageous for some optical applications, here it essentially makes painstaking spectral matching of single QDs and individual PCR modes (second factor in the r.h.s of Eq.\,\ref{equ:FP}) dispensable.  However, LDOS mapping will still be possible under experimental conditions with reduced spectral diffusion, if a sufficient number of cavities with varying geometry parameters is provided. Under these conditions, fine tuning of the spectral overlap between QD emitter and any given cavity mode will become possible with high resolution, as demonstrated in principle in Figs.\,\ref{fig:4}(b) and \ref{fig:4}(c). 

The size of the Ge QDs impedes to some extent the probing accuracy of narrow LDOS features, such as the minima or maxima of modes M2 and M3. Also, there remains some ambiguity regarding the absolute LDOS minima, which are partly obscured by the aforementioned background signal caused by the Ge wetting layer (Fig.\,\ref{fig:5}(d)). The main discrepancies between experiments and simulations occur for modes that have their LDOS maximum at the very edges of the cavity, in particular modes M1 and M3 (Fig.\,\ref{fig:mode-pattern}).  The edge position is in fact a critical location for a QD, as can be seen in Fig.\,\ref{fig:SEM}(b). Because of its finite size, parts of the QD are removed by air-hole etching. Such a damaged QD with one flank terminated by the non-passivated surface of the air hole cannot be assumed to behave in the same way as an emitter further inside the cavity that is fully embedded in the Si matrix. The reduction of $\tau_{nr}$ for such truncated edge-QDs leads to $I_{PL}$ significantly smaller than expected from the simulated LDOS at this position

\section{Summary}
\label{sec:Sum}

In this work we demonstrated the outstanding potential of deterministically positioned single Ge QDs for LDOS mapping in 2D PCRs. Compared to other approaches Ge QDs have several distinct advantages for this type of application: (i) Site-control has reached an unrivaled state of perfection in the Ge/Si heterosystem, with reproducible positioning accuracies of individual Ge QDs better than 20\,nm having been achieved. (ii) Our LDOS mapping results clearly demonstrate the ability of precise and deterministic matching of SiGe QD position and absolute LDOS maximum of PCR modes, which is essential for a vast number of quantum optical applications. Matching of dot position and LDOS maximum of an arbitrary PCR is performed in a single process run only requiring the alignment of the two mask layers for the PC structures and the assigned QDs. Additionally, our method for the site control of Ge QDs is compatible with standard Si device technology and, thus, allows for parallel processing and scalability, a prerequisite for a potential implementation in a Si based integrated optics platform.


\section*{Funding Information}
Austrian Science Fund (FWF) (F2502-N17, F2512-N17 of the Spezialforschungsbereich IRON (SFB025); P29137-N36).

\section*{Acknowledgments}

We thank G. Katsaros for providing the custom-made, re-bonded SOI substrates. We are grateful to R. Jannesari, M. Grydlik and D. Gerace for valuable discussions as well as to A. Halilovi\'c and S. Br\"auer for technical clean-room support during the fabrication of the PCRs.

\bibliography{/home/thomas/Documents/paper_poster/Schatzl_Ordered_QD_in_PhC/LatexVersion/References_L3_mapping_Flo_corr}

\end{document}